\documentclass[aps,twocolumn,superscriptaddress,floatfix,nofootinbib]{revtex4-1}
\usepackage{graphicx,amsmath,amssymb,verbatim,color}
\usepackage{booktabs}
\usepackage{comment}
\usepackage{soul}
\usepackage[dvipsnames]{xcolor}
\usepackage{bm}
\usepackage[utf8]{inputenc}
\usepackage[colorlinks=true,citecolor=blue,linkcolor=blue,urlcolor=blue, backref=false,pdfborder={0 0 0}]{hyperref}
\usepackage{float}
\usepackage{multirow}
\newcommand{\be}{\begin{equation}\begin{gathered}}
\newcommand{\ee}{\end{gathered}\end{equation}} 
\newcommand{\barr}{\begin{eqnarray}}
\newcommand{\earr}{\end{eqnarray}} 
\usepackage{romannum}
\usepackage[normalem]{ulem}

\def\hinvMpc{h\,{\rm Mpc}^{-1}}

\begin{document}

\pagenumbering{arabic}

\title{
Cosmological constraints from combined probes \\with the three-point statistics of galaxies at one-loop precision
}
\author{Simon Spaar}
\affiliation{Institute for Theoretical Physics, ETH Zürich, 8093 Zürich, Switzerland} 
\author{Pierre Zhang}
\affiliation{Institute for Theoretical Physics, ETH Zürich, 8093 Zürich, Switzerland} 
\affiliation{Institute for Particle Physics and Astrophysics, ETH Zürich, 8093 Zürich, Switzerland}

\date{\today} 
\begin{abstract}
We present cosmological constraints from a joint analysis including the power spectrum and bispectrum of BOSS galaxies based on the Effective Field Theory of Large-Scale Structure predictions at one-loop order, in combination with CMB data from \textit{Planck}, Supernovae from Pantheon+, and BAO from eBOSS and 6dF/MGS. 
Limits on $\Lambda$CDM parameters are in good agreement, and on average $\sim 5-10\%$ tighter, compared to former results including similar datasets but no bispectrum. 
Moreover, we find that galaxies at the three-point level with one-loop precision are decisive for the dark energy equation of state, constrained to be $w =-0.975 \pm 0.019$ at $68\%$CL.  
This value, consistent at $\sim 1.3\sigma$ with a cosmological constant, represents an improvement of about $140\%$ with respect to former determination. 
Our analyses illustrate the importance of beyond-two-point statistics at the highest reachable scales in constraining cosmological parameters, and in particular departure from $\Lambda$CDM. 
\end{abstract}

\maketitle

\section{\label{sec:intro}Introduction}
The distribution of matter in the early Universe is Gaussian, yet the distribution of galaxies at late times is non-Gaussian. 
Gravitational collapse leads to the formation of a rich variety of structures that we see in galaxy surveys. 
If we are interested in describing the large scales, we are then facing two complications: 
First, large scales receive contributions from mode-coupling to nonlinear small-scale features. 
Second, the galaxy distribution is no more simply described by the two-point correlation function, as nonlinear contributions arise in higher $N$-point functions. 
To extract optimally cosmological information from the statistics of galaxies, one is thus faced with the challenge of describing the nonlinear dynamics of galaxies under gravity on the one hand, and, on the other hand, the analysis of $N$-point functions beyond the two-point level. 

Estimating $N$-point functions from galaxy surveys, especially beyond the two-point, is not a trivial task. 
Thanks to an intensive line of research, there is a rather long history of cosmological analyses of the two-point and three-point functions of galaxies, or their Fourier counterparts, respectively the power spectrum and bispectrum (see e.g.,~\cite{Gil-Marin:2016wya} and Refs. therein). 
As suggested above, one other aspect that complicates their treatment is their modeling beyond the linear scales, which limits severely the information we can retrieve. 
As such, until recently, in galaxy surveys most of the cosmology was extracted through the imprints of the baryon acoustic oscillations (BAO), or relying on approximate modeling to infer the scale-independent amplitude $f\sigma_8$ from the redshift-space distortions (see e.g.,~\cite{BOSS:2016wmc,eBOSS:2020yzd}). 

The situation has changed in the last decade, due to several aspects. 
On the one hand, the quality of spectroscopic observations and their data processing has reached with the Baryon Oscillation Spectroscopic Survey (BOSS) of the Sloan Digital Sky Survey (SDSS)~\cite{BOSS:2016wmc} a level that aligns with standards of the era of precision cosmology we have entered. 
Thanks to exquisite BAO/$f\sigma_8$ measurements, combinations with Cosmic Microwave Background (CMB) data from the \textit{Planck} satellite and supernovae observations from Pantheon have lead to the most precise determination of the cosmological concordance model, $\Lambda$CDM, and its canonical extensions~\cite{BOSS:2016wmc,Planck:2018vyg,eBOSS:2020yzd,Brout:2022vxf}. 
On the other hand, building on significant prior research (see e.g.,~\cite{Bernardeau:2001qr} for a review) a consistent treatment of the gravitational collapse of structures at long distances has emerged within the framework of Effective-Field Theory of Large-Scale Structure (EFTofLSS)~\cite{Baumann:2010tm,Carrasco:2012cv}. 
As an EFT approach, it allows for a systematic organization at long distances of the fluid expansion of the density and velocity fields of galaxies in perturbations and gradients, integrated over their whole history~\cite{Senatore:2014eva}. 
Including basically all terms allowed by the equivalence principle, the predictions incorporate all relevant aspects entering the description of galaxies at long distances: small-scales physics~\cite{Carrasco:2012cv,Carrasco:2013mua}, galaxy biasing~\cite{McDonald:2006mx,McDonald:2009dh,Assassi:2014fva,Senatore:2014eva,Mirbabayi:2014zca,Angulo:2015eqa}, spatial gradients~\cite{Fujita:2016dne}, redshift-space distortions~\cite{Senatore:2014vja,Lewandowski:2015ziq,Perko:2016puo}, long-wavelength displacements around the BAO peak~\cite{Matsubara:2007wj,Matsubara:2008wx,Porto:2013qua,Senatore:2014via,Baldauf:2015xfa,Vlah:2015sea,Vlah:2015zda,Senatore:2017pbn}, baryons~\cite{Lewandowski:2014rca,Schmidt:2016coo,Braganca:2020nhv}, or massive neutrinos~\cite{Blas:2014hya,Senatore:2017hyk,Garny:2020ilv}. 
What's more, non-standard assumptions can be systematically incorporated from first principles, as for examples primordial non-Gaussianities~\cite{Baldauf:2010vn,Assassi:2015fma,Lewandowski:2015ziq}, modified gravity~\cite{Lewandowski:2016yce,Cusin:2017wjg} or dark long-range interactions~\cite{Bottaro:2023wkd}. 
Besides, efficient evaluation schemes have been developed to compute the nonlinear corrections and IR-resummation~\cite{Schmittfull:2016jsw,Simonovic:2017mhp,Ivanov:2018gjr,Lewandowski:2018ywf,DAmico:2020kxu}, enabling cosmological inference with the full shape of galaxy statistics based on EFT predictions. 
Hand in hand, full-shape analyses at the two-point level have lead to CMB-independent determinations of $\Lambda$CDM parameters from BOSS/eBOSS data~\cite{DAmico:2019fhj,Ivanov:2019pdj,Colas:2019ret,Troster:2019ean,Philcox:2020vvt,Chen:2021wdi,Zhang:2021yna,DAmico:2021ymi,Ivanov:2021fbu,Zhang:2021uyp,Chen:2022jzq,Simon:2022lde,Ivanov:2021zmi,Simon:2022csv,Chudaykin:2022nru,Donald-McCann:2023kpx,Zhao:2023ebp,Ramirez:2023ads}. 
Canonical extensions or more exotic hypotheses, in combination with other probes, have also been explored using the full shape based on the EFT predictions (see e.g.,~\cite{Ivanov:2019hqk,DAmico:2020kxu,DAmico:2020tty,Chudaykin:2020ghx,Piga:2022mge,Ballardini:2023mzm,Carrilho:2022mon,Lague:2021frh,Rogers:2023ezo,Gonzalez:2020fdy,Allali:2021azp,Allali:2023zbi,Schoneberg:2023rnx,Simon:2022ftd,Simon:2022csv,Rubira:2022xhb,Niedermann:2020qbw,Cruz:2023cxy,DAmico:2020ods,Ivanov:2020ril,Smith:2020rxx,Herold:2021ksg,Reeves:2022aoi,Simon:2022adh,Gsponer:2023wpm,Bottaro:2023wkd,Camarena:2023cku,He:2023oke}). 
Dedicated tools for fast evaluation of EFTofLSS statistics have been developed independently~\cite{McEwen:2016fjn,DAmico:2020kxu,Chudaykin:2020aoj,Chen:2020fxs,Chen:2020zjt,Noriega:2022nhf}, providing complementary ways to analyze the data of tomorrow, as currently assessed in DESI~\cite{Aghamousa:2016zmz} or Euclid~\cite{Amendola:2016saw,Euclid:2023bgs}. 

Until now, most of the theoretical developments and analysis methods have focused on the two-point function. 
The inclusion of the first nonlinear corrections, of the so-called one-loop order, increases the $k$-reach of the predictions by roughly a factor $2$ to $3$~\cite{DAmico:2019fhj,DAmico:2021ymi}. 
This holds the promises to enhance greatly our access to the cosmological information, as at first sight it seems that the one loop leads to a factor of $\sim 10$ in the number of modes probed. 
There are two caveats to this naive view we want to point out. 
First, the inclusion of nonlinear corrections comes with the price of adding a multitude of EFT parameters to model all possible responses of the gravitational potential (and velocity divergence) to galaxies, small-scale physics, stochasticity, baryons, and so on. 
Thus, parts of the information are lost when marginalizing over those nuisances. 
Second, as we stressed right at the start, the Universe is non-Gaussian, even at large scales.  
In fact, with current-stage surveys, it was shown that once combined with \textit{Planck}, the constraints were only marginally improving, if at all, with respect to \textit{Planck} + BAO/$f\sigma_8$ results~\cite{Ivanov:2019hqk,DAmico:2020kxu,Simon:2022csv}. 
Based on these observations, it is then natural to consider higher-$N$-point functions to better determine the EFT parameters, mitigating the information loss, and second to retrieve the information residing in non-Gaussian statistics. 
In order for this to occur, however, it is crucial to consider them at the shortest possible scales. 
The deeper we dive, the more nonlinear the Universe appears. 
In perturbation theory, it means that as we approach the nonlinear scale (or EFT breakdown scale) all corrections become order one with respect to the linear contribution, and so does the importance of higher-$N$-point functions with respect to the two-point. 
As a matter of fact, at tree level, the bispectrum is bringing only mild improvements with respect to the results from the one-loop power spectrum only, of about $\sim 10\%$ with BOSS data~\cite{DAmico:2019fhj,Ivanov:2021kcd,Philcox:2021kcw}. 
Clearly, there is a call to go beyond. 

Recently, a new step has been made in this direction: 
the bispectrum of BOSS galaxies has been analyzed at one-loop precision~\cite{DAmico:2022osl}. 
This analysis has lead to an uncertainty reduction of about $30\%, 18\%$, and $13\%$ on $\sigma_8$, $H_0$, and $\Omega_m$, respectively, compared to the analysis with the power spectrum only. 
Naively, reducing an uncertainty $\sigma$ by $\sim 30\%$ corresponds to a doubling in size of the data volume $V$, as $\sigma \propto V^{-1/2}$.  
Two important ingredients were necessary to make this possible. 
First, the theory of galaxies in redshift space had to be extended to compute the one-loop corrections to the bispectrum with appropriate counterterms to absorb the UV-sensitivity of the loops. 
This was done in Ref.~\cite{DAmico:2022ukl} (see also Refs. therein), where in particular subtleties in the renormalization of the momentum operator and velocity products involved in redshift space had to be elucidated. 
Second, an efficient algorithm had to be developed in order to make the evaluation of the bispectrum one-loop predictions fast enough for cosmological inference. 
This was provided in Ref.~\cite{Anastasiou:2022udy}.  
There, the proposed algorithm relies on the decomposition of the input linear power spectrum onto a limited number of fitting functions such that loop integrals appear then like massive propagators that can be computed analytically using techniques borrowed from quantum chromodynamics (QCD). 
The upshot is that evaluating loop integrals then boils down to simple tensor multiplications with small enough dimension to be efficient. 
All in all, we now have at hand a fantastic new tool to extract the cosmological information from the galaxy maps. 
In Ref.~\cite{DAmico:2022gki}, the one-loop bispectrum of galaxies was used to set limits on primordial non-Gaussianities, in particular beyond the local type (see also~\cite{Cabass:2022wjy,Cabass:2022ymb} for limits obtained with the tree-level bispectrum), opening a new window onto the Universe's first instants.  

Ultimately, we are interested in the cosmology inferred from a joint analysis of all datasets (consistent enough to be combined). 
In this work, we perform the first multi-probe analysis including the galaxy three-point statistics at one-loop precision. 
In light of cosmological observations, we analyze two models: $\Lambda$CDM and $w$CDM, where in the latter the cosmological constant $\Lambda$ is replaced by a dark energy component with equation of state $w$. 
Our paper is organized as follow: 
In Sec.~\ref{sec:setup}, we present the datasets, likelihoods, and methods used in our cosmological analyses, for which the results are presented in Sec.~\ref{sec:results}. 
We conclude and provide final remarks in Sec.~\ref{sec:conclusion}. 
In App.~\ref{app:triangle_plots}, we provide the full cosmological triangle plots of the posteriors from our analyses. 

\subsection{\label{sec:estimate} The higher, the better}
One may wonder how informative is the bispectrum of galaxies at one-loop precision. 
Before performing the data analysis, we want to highlight the importance to include higher modes \emph{especially} when considering higher $N$-point functions. 
As sketched above, the closer we get to the nonlinear scale $1/k_{\rm NL}$, the closer the relative density becomes $\sim \mathcal{O}(1)$, and so does the relevance of all $N$-point functions. 
To be slightly more quantitative, we can compare the signal-to-noise (SNR) of the bispectrum $B$ to the power spectrum $P$, 
\begin{equation}
r \equiv \frac{\textrm{SNR}(B)}{\textrm{SNR}(P)} \ , 
\end{equation}
where $\textrm{SNR}(P) = P \cdot C_P^{-1} \cdot P$ and $\textrm{SNR}(B) = B \cdot C_B^{-1} \cdot B$. 
When including modes up to $k_{\rm max}^B$ for the bispectrum and up to $k_{\rm max}^P$ for the power spectrum, a simple estimate is given by
\begin{equation}\label{eq:snr}
r \simeq 12\pi \left(\frac{k_{\rm max}^B}{k_{\rm NL}}\right)^{3 + n} \left(\frac{k_{\rm max}^B}{k_{\rm max}^P}\right)^{3}\ ,
\end{equation}
where $n \sim -1.5$ is the $\log$-slope of the power spectrum around $k_{\rm max}^B$. 
To arrive at this, we assume Gaussian covariances,
\begin{equation}
C_P \simeq \frac{2}{V} \frac{(2\pi)^3}{N_k} P P \ , \quad C_B \simeq \frac{6}{V} \frac{(2\pi)^6}{N_T} P P P \ ,
\end{equation}
where $V$ is the survey volume. 
For simplicity, we take $N_k \simeq 4 \pi (k_{\rm max}^P/2)^2 k_{\rm max}^P$ and $N_T \simeq 8\pi^2 (k_{\rm max}^B/2)^3 (k_{\rm max}^B)^3$. 
Next, we estimate the bispectrum size as
\begin{equation}
B(k) \sim 6 \, P(k) P(k)\ ,
\end{equation}
working in the approximation where all momenta are of the same order. 
As such, we obtain
\begin{equation}
r \simeq \frac{12\pi}{ (2\pi)^3} \frac{(k_{\rm max}^B)^6}{(k_{\rm max}^P)^3} \  P(k_{\rm max}^B) \ ,
\end{equation}
and hence Eq.~\eqref{eq:snr}, taking further 
\begin{equation}
P(k) \simeq (2\pi)^3 \left( \frac{1}{k_{\rm NL}}\right)^3 \left(\frac{k}{k_{\rm NL}}\right)^n \ .
\end{equation}
It is clear from our estimated relative SNR that the bispectrum gains increasing importance compared to the power spectrum as we access more scales. 
Eq.~\eqref{eq:snr} illustrates that, near the nonlinear scale, there is a growing number of triple counts in comparison to pairs while the density reaches order one. 
Approximately, at tree level, $k_{\rm max}/k_{\rm NL} \sim 0.2$, while at one loop, $k_{\rm max}/k_{\rm NL} \sim 0.5$. 
Thus, compared the power spectrum at one loop, $r\sim 0.2$ or $r\sim 13$ when considering the bispectrum at tree level or one loop, respectively. 
These relative SNR values can be translated into a reduction on uncertainties from the inclusion of the bispectrum by approximately $10\%$ and $70\%$, respectively. 
This reduction is determined considering that $\sigma_{PB} \simeq \sigma_P / \sqrt{ 1 + \sigma_P^2/\sigma_B^2 } \simeq \sigma_P / \sqrt{ 1 + r }$, where we proxy $1/\sigma_P^2 \sim \textrm{SNR}(P)$ and $1/\sigma_B^2 \sim \textrm{SNR}(B)$. 
It thus appears that the three-point function of galaxies carries important additional information beyond the two-point especially when analyzed at the one loop. 
Keeping in mind that these are rough estimates, which e.g., do not account for parameter degeneracies, we now turn to the data analysis.

\section{\label{sec:setup}Methodology}
We perform multi-probe analyses of $\Lambda$CDM and $w$CDM including for the first time both power spectrum and bispectrum of galaxies at one-loop precision. 

\subsection{\label{sec:data}Datasets}
We carry out various analyses from a combination of the following datasets:
\begin{itemize}
    \item {\bf BOSS:} The SDSS-III BOSS DR12 galaxy sample data~\cite{BOSS:2016wmc}. 
    We make use of the full shape of the power spectrum and bispectrum, referred respectively as {\bf 2pt} and {\bf 3pt}. 
    Descriptions of the measurements and covariances can be found in Ref.~\cite{DAmico:2022osl}. 
    In order to gauge the additional information brought by the full-shape statistics at one loop, we sometimes compare with results obtained using from BOSS only the BAO parameters, referred as {\bf BAO}~\cite{BOSS:2016wmc}. 
    Notice that the BOSS BAO parameters are obtained on reconstructed measurements, therefore including the BAO information from the 2pt and partially from the 3pt and higher-$N$-point functions. 
    \item {\bf Planck:} Cosmic microwave background temperature and polarization anisotropies through the \textit{Planck} likelihood of high-$\ell$ TT, TE, EE + low-$\ell$ TT, EE + lensing~\cite{Planck:2018vyg}. 
    Small correlation in the lensing and integrated Sachs-Wolfe effects with clustering data are neglected. 
    \item {\bf ext-BAO:} 
    Joint BAO constraints from eBOSS DR14 Lyman-$\alpha$ absorption auto-correlation at $z = 2.34$ and cross-correlation with quasars at $z = 2.35$~\cite{Agathe:2019vsu,Blomqvist:2019rah}, together with BAO measurements from 6dFGS at $z = 0.106$ and SDSS DR7 MGS at $z = 0.15$~\cite{Beutler:2011hx,Ross:2014qpa}. 
    \item {\bf Pantheon+:} The Pantheon+ catalog of uncalibrated luminosity distance of type Ia supernovae (SNIa) in the range $0.01 < z < 2.26$~\cite{Brout:2022vxf}. 
\end{itemize}

We consider the following combinations: 
\begin{itemize}
    \item {\bf base}: All cosmological probes considered in this work other than BOSS galaxies ({\bf Planck + ext-BAO + Pantheon+}), making for the baseline multi-probe for comparison. 
    \item {\bf base+2pt+3pt}: Primary multi-probe analysis including both the power spectrum and bispectrum of BOSS galaxies at one-loop precision. 
    \item {\bf base+2pt}: To assess the constraining power brought by the galaxy three-point statistics, we remove 3pt from the primary analysis. 
    \item {\bf base+BAO}: To assess the information brought by the one-loop precision analysis of galaxy statistics, we exclude all full-shape galaxy data while retaining the standard BAO measurements (in particular the one of BOSS). 
\end{itemize}

\subsection{\label{sec:likelihood}Galaxy 1-loop 2pt+3pt likelihood}
We analyze the full shape of the power spectrum (2pt) and bispectrum (3pt) of BOSS galaxies in redshift space using the one-loop (1-loop) predictions based on the EFTofLSS presented in Refs.~\cite{Perko:2016puo}~and~\cite{DAmico:2022ukl}, respectively. 
Our predictions include all necessary nonlinear corrections at the one-loop order: perturbation theory contributions of biased tracers in redshift space, counterterms / stochastic terms, with the latter shown to be sufficient and necessary to renormalize the former. 
We account for long-wavelength displacements around the BAO peak using the full Lagrangian resummation for the power spectrum~\cite{DAmico:2020kxu} and a simplified scheme relying on a wiggle-no wiggle split of the input linear power spectrum for the bispectrum shown to be accurate enough for BOSS volume~\cite{DAmico:2022osl}. 
Additional modeling such as Alcock-Paczynski distortions, window function, and binning, have been extensively tested and are described in Refs.~\cite{DAmico:2019fhj}~and~\cite{DAmico:2022osl} for the power spectrum and bispectrum, respectively. 
Thanks to our one-loop predictions including a consistent treatment in redshift space of nonlinearities, imprints of the BAO, and observational modeling, modes in galaxy surveys beyond the linear regime can be reached with a controlled theory error and with good fit to data. 
Extensive tests on the power spectrum and its $k$-reach have been performed against high-fidelity simulations or based on perturbative arguments in Refs.~\cite{Nishimichi:2020tvu,DAmico:2019fhj,Chen:2020zjt,DAmico:2021ymi,Zhang:2021uyp}, and in Ref.~\cite{DAmico:2022osl} for the bispectrum. 

The likelihood we use in this work is described in Ref.~\cite{DAmico:2022osl}. 
Following the BOSS survey specifications, the data are divided in two redshift bins $0.2<z<0.43 \  (z_{\rm eff}=0.32)$, $0.43<z<0.7  \ (z_{\rm eff}=0.57)$, where for each, we consider a north cut and south cut respective to the galactic plane, for a total of four skies. 
For each sky, we make use of the monopole and quadrupole of both the power spectrum and bispectrum. 
With the exception of the bispectrum quadrupole, we fit them using the one-loop predictions. 
For those we include scales between $k_{\rm min}=0.01 \hinvMpc$ to avoid spurious effects from unmodeled large-scale systematics and $k_{\rm max} = 0.20/0.23 \hinvMpc$ for the low-$z$ and high-$z$ skies, respectively, based on previous estimations of the theory error with respect to BOSS uncertainties~\cite{Colas:2019ret,Zhang:2021yna,DAmico:2022osl}. 
For the bispectrum quadrupole that we fit with tree-level predictions, we include scales only up to $k_{\rm max} = 0.08 \hinvMpc$ for all skies.

For each sky, we construct a data vector $D_\alpha$ of all observables concatenated, where $\alpha$ is a generic index that runs on multipoles, the $k$-bins of the power spectrum, or the triangle bins of the bispectrum. 
Furthermore, we estimate the covariance $C_{\alpha \beta}$ from the scatter across the corresponding measurements of the $2048$ patchy mocks~\cite{Kitaura:2015uqa}, and correct the inverse covariance matrix by the Hartlap factor~\cite{Hartlap:2006kj}. 
Then, at each likelihood evaluation, we compute the corresponding theory vector $T_\alpha(\theta)$ where $\theta$ is the parameter vector composed of the cosmological parameters of interest and the EFT parameters entering in the predictions of the galaxy statistics. 
For each sky, the likelihood $\mathcal{L}(D_\alpha | \theta)$ then reads: 
\begin{equation}
-2 \ln \mathcal{L} = \sum_{\alpha, \beta} (D_\alpha - T_\alpha(\theta)) \cdot C^{-1}_{\alpha \beta} \cdot (D_\beta - T_\beta(\theta)) \ . 
\end{equation}
For the predictions to be valid within perturbation theory, the EFT parameters are expected to be $\sim \mathcal{O}(1)$. 
We therefore further marginalize over them with a Gaussian prior centered on $0$ of width $\sim 2$ to keep them within physical range, with the exception of $b_1$, that is always positive, for which we use an equivalent log-normal prior. 
Moreover, we account for the fact that the expected EFT parameters differ between skies. 
This discrepancy arises from the effects of redshift evolution on the one hand and small variations in observations of the north and south galactic hemisphere on the other hand. We therefore assign one set of EFT parameters per sky while imposing correlations between them across the skies.
For each EFT parameter $b_i$ coming in a quadruplet for the four skies, we define a prior such that we expect the values $b_i$ to be different only by $10\%$ between the north and south skies (within a redshift bin), and by $20\%$ between the low-$z$ and high-$z$ redshift bins (within a hemisphere). 
In practice, this generalizes the Gaussian prior previously described to a multivariate Gaussian prior for a given quadruplet. 
Technically, as our primary focus in this work are cosmological parameters, we analytically marginalize over the EFT parameters that enter linearly in the predictions, and thus quadratically in the likelihood, using properties of Gaussian integrals. 
This makes our analysis computational tractable, as we then only need to scan over $3$ EFT parameters (the galaxy biases contributing to tree-level predictions, i.e., $b_1$, $b_2$, and $b_5$ in the notation of Ref.~\cite{DAmico:2022ukl}) out of total of $41$ EFT parameters per sky. 
See Ref.~\cite{DAmico:2022osl} for more details regarding priors and the marginalization procedure.

\subsection{Inference setup}

For all analyses, while marginalizing over nuisance parameters of each likelihood, we scan the $\Lambda$CDM parameters within large uniform priors, i.e., $\{ \omega_b, \omega_{\rm cdm}, H_0, \ln (10^{10}A_{s }), n_s \}$, respectively the baryons abundance, the cold dark matter abundance, the Hubble constant, the log-amplitude of the primordial fluctuations, and the spectral tilt. 
Since we are using the \textit{Planck} likelihood, we additionally scan over the optical depth to reionization, $\tau_{reio}$. 
We also present our results for derived parameters $\Omega_m, \sigma_8$, and $S_8$, respectively the fractional matter abundance, the clustering amplitude, and the lensing amplitude. 
Following \textit{Planck} prescription, we consider two massless neutrinos and one massive neutrino fixed to minimal mass, $\sum m_\nu = 0.06 \ e{\rm V}$~\cite{Planck:2018vyg}. 

Additionally, we explore an extension to $\Lambda$CDM where dark energy is assumed to be a generalized dynamical fluid with equation of state $w$, replacing the cosmological constant. 
Taking a data-driven approach, in first place we let $w$ vary freely within a large uniform prior. Especially, $w \neq -1$ would signal a departure from $\Lambda$. 
Alternatively, we restrict $w \geq -1$, the physical region of a broad class of dark energy models as we discuss below. 

Posterior distributions are sampled using the Metropolis-Hastings algorithm from the cosmological inference software \texttt{MontePython-v3}~\cite{Audren:2012wb,Brinckmann:2018cvx}~\footnote{\url{https://github.com/brinckmann/montepython_public}} interfaced with the Boltzmann code \texttt{CLASS}~\cite{Blas:2011rf}~\footnote{\url{http://class-code.net}} for the linear cosmology, and with the code~\texttt{PyBird}~\cite{DAmico:2020kxu}~\footnote{\url{https://github.com/pierrexyz/pybird}} for the nonlinear `one-loop' cosmology and related likelihoods.  
All chains presented in this paper are converged according to Gelman-Rubin criterion $R-1 \lesssim 0.01$.
Plots and credible intervals of marginalized posteriors are obtained using \texttt{GetDist}~\cite{Lewis:2019xzd}. 
Best fits and $\chi^2$ are obtained minimizing using \texttt{iminuit}~\cite{iminuit} and further refined following the procedure outlined in appendix of Ref.~\cite{Schoneberg:2021qvd}. 

Note that while our marginalized posteriors are in principle subject to projection effects, here we are using strong combinations of datasets such that these effects are not important, as discussed in Ref.~\cite{Simon:2022lde}. 
For example, we have checked that the best fits always lie well within the $68\%$ credible intervals. 

\section{\label{sec:results}Results}
\subsection{$\Lambda$CDM}

\begin{figure}[ht]
\includegraphics[width=.48\textwidth]{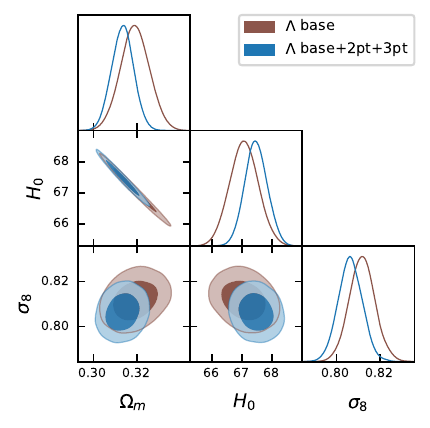}
\caption{$\Lambda$CDM triangle plot of $\Omega_m$, $H_0$, and $\sigma_8$ from our base combined probes and in combination with BOSS 1-loop 2pt + 3pt. }
\label{fig:lcdm_triangle_plot_small}
\end{figure}

\begin{table*}[]
    \centering
    \begin{tabular}{|l|c|c|c|c|}
    \hline
    \multicolumn{5}{|c|}{$\Lambda$CDM} \\
    \hline
        & base &  base+BAO & base+2pt &  base+2pt+3pt \\
    \hline
$10^{2}\omega_{b }$
	& $2.230\pm 0.014$
        & $2.237\pm 0.013$
        & $2.234\pm 0.014$
        & $2.235\pm 0.013$
    \\
$\omega_{\rm cdm }$
        & $0.1206\pm 0.0011$
        & $0.11967\pm 0.00089$
        & $0.11996\pm 0.00096$
        & $0.11967\pm 0.00084$
    \\
$H_0 \ [\textrm{km/s/Mpc}]$
        & $67.07\pm 0.48$
        & $67.48\pm 0.40$
        & $67.35\pm 0.43$
        & $67.47\pm 0.38$
    \\
$\ln(10^{10}A_{s })$
        & $3.044\pm 0.014$
        & $3.048\pm 0.014$
        & $3.046\pm 0.014$
        & $3.037\pm 0.014$
    \\ 
$n_s$
        & $0.9628\pm 0.0039$
        & $0.9652\pm 0.0036$
        & $0.9643\pm 0.0037$
        & $0.9646\pm 0.0035$
    \\
$\tau_{\rm reio }$
        & $0.0539\pm 0.0073$
        & $0.0564^{+0.0067}_{-0.0074}$
        & $0.0553\pm 0.0073$
        & $0.0518\pm 0.0069$
    \\
    \hline
$\Omega_{m }$
        & $0.3192\pm 0.0067$
        & $0.3134\pm 0.0054$
        & $0.3152\pm 0.0059$
        & $0.3135\pm 0.0051$
    \\
$\sigma_8$
        & $0.8117\pm 0.0059$
        & $0.8105\pm 0.0060$
        & $0.8107\pm 0.0060$
        & $0.8063\pm 0.0056$
    \\
$S_8$
        & $0.837\pm 0.012$
        & $0.828\pm 0.010$
        & $0.831\pm 0.011$
        & $0.8242\pm 0.0095$
    \\
    \hline
    \end{tabular}
    \caption{$\Lambda$CDM $68\%$ credible intervals from the various dataset combinations considered in this work.  }
    \label{tab:lcdm}
\end{table*}

We now present our $\Lambda$CDM results obtained from the various dataset combinations considered in this work. 
In Fig.~\ref{fig:lcdm_triangle_plot_small}, we show the triangle plot of $\Omega_m, H_0$, and $\sigma_8$ while the full cosmological triangle plot is provided in App.~\ref{app:triangle_plots}. 
The corresponding $68\%$ credible intervals are given in Tab.~\ref{tab:lcdm}. 
First, we note that for all cosmological parameters, the constraints from the various combinations are consistent within $1\sigma$. 
Compared to the baseline multi-probe, the inclusion of BOSS 1-loop 2pt+3pt reduces the uncertainties on $\omega_{\rm cdm}, \Omega_m, H_0, S_8$ by about $20-25\%$, $\omega_b, n_s$ by $\sim 10\%$, and $\sigma_8, \tau_{\rm reio}$ by $\sim 5\%$. 
The error bars on $A_s$ are similar. 
Notably, our full combination leads to constraints that are tighter than the ones from `base+BAO', although mildly, by about $4-7\%$ for all cosmological parameters but $\omega_b$ and $A_s$, for which the error bars are similar. 
This stands in contrast to the limited improvement from BOSS 1-loop 2pt, which actually is less informative than the reconstructed BAO when combined with the baseline probes. 
In fact, Refs.~\cite{Ivanov:2019hqk,DAmico:2020kxu} showed that even including reconstructed BAO through cross-correlation with the BOSS 1-loop 2pt does not enhance the constraints over using BAO information alone when jointly fitting with \textit{Planck}. 
This situation is largely attributed to the current volume of galaxy data compared to CMB data. 
The full shape of the two-point function contains further information beyond BAO, as shown in isolation in e.g.,~\cite{Philcox:2020xbv,Smith:2022iax}. 
In summary, galaxy data with one-loop predictions can significantly impact $\Lambda$CDM constraints when combining all cosmological observations, as displayed when including the bispectrum. 

As for the mean values, the inclusion of BOSS 1-loop 2pt+3pt pulls $\Omega_m, H_0$, $\sigma_8$ towards values inferred when analyzed alone~\cite{DAmico:2022osl}. 
Interestingly, $S_8$ from our full combination is in better agreement with weak lensing measurements. 
Quoting $S_8 = 0.775 \pm 0.025$ inferred from DES Y3 3x2pt~\cite{DES:2021wwk}, the $\sigma$-deviation (assuming Gaussian posteriors) from the baseline multi-probe is $2.2\sigma$, whereas with the inclusion of BOSS 1-loop 2pt+3pt, it reduces to $1.8\sigma$. 
Compared to the joint cosmic shear analysis of DES Y3 and KIDS-1000~\cite{Kilo-DegreeSurvey:2023gfr}, which yielded an estimate of $S_8 = 0.790 ^{+0.018}_{-0.014}$, our full combination agrees at a $\sim 1.7\sigma$ level. 
This agreement holds even when using a straightforward Gaussian estimate and without accounting for projection effects in the former (for further discussion, see Ref.~\cite{Kilo-DegreeSurvey:2023gfr}). 

To assess the impact of BOSS data on the fit, we evaluate the cost in $\chi^2$ at its minimum for the baseline multi-probe when considering either the inclusion of BOSS BAO, 2pt, or 2pt+3pt. 
The resulting values are $\Delta \chi^2 \simeq 0.3, 10.3, 11.4$, respectively. 
Correspondingly, the $\chi^2$ of {\it Planck} is modified by an amount of $\Delta \chi^2 \simeq 0.3, 9.4, -2.6$. 
We can make the following observations. 
The addition of BOSS 2pt results in a notable increase in $\chi^2$ of $\sim 10$ in the fit to the baseline multi-probe. 
The majority of this difference, $\Delta \chi^2 \sim 9$, is observed in the fit to {\it Planck}. 
In contrast, the further inclusion of BOSS 3pt has a relatively minor impact on the fit to the baseline probes as a whole, with $\Delta \chi^2 \sim 1$. 
Perhaps more interestingly, the inclusion of BOSS 3pt actually improves the fit to {\it Planck}, reducing the $\chi^2$ by approximately $-2.6$, compared to the fit to the baseline multi-probe. 
At the same time, the fit to Pantheon+ is degraded by $\Delta \chi^2 \sim 14$. 
At face value, it appears that our strongest combination of BOSS data, 2pt+3pt, is in good agreement with {\it Planck}, while Pantheon+ seems to be in slight tension with both {\it Planck} and BOSS. 
This can be partially attributed to the relatively high value of $\Omega_m = 0.338 \pm 0.018$ favored by Pantheon+~\cite{Brout:2022vxf}, which contrasts with the preferred values of $\Omega_m = 0.315 \pm 0.007$ from {\it Planck}~\cite{Planck:2018vyg} or $\Omega_m = 0.311 \pm 0.010$ from BOSS 2pt+3pt~\cite{DAmico:2022osl}. 
Consequently, the lower value of $S_8$ favored by our full combination can be mainly attributed to its correlation with $\Omega_m$, though a slight shift in the amplitude of the primordial fluctuations $A_s$ is also observed. See Fig.~\ref{fig:lcdm_triangle_plot_full} for a visual representation. 

In the future, it would be interesting to perform our analysis in light of the updated likelihood of \textit{Planck} final data release~\cite{Tristram:2023haj}, where $S_8$ is in better agreement with weak lensing.

\subsection{$w$CDM}

\begin{figure}[ht]
\includegraphics[width=0.47\textwidth]{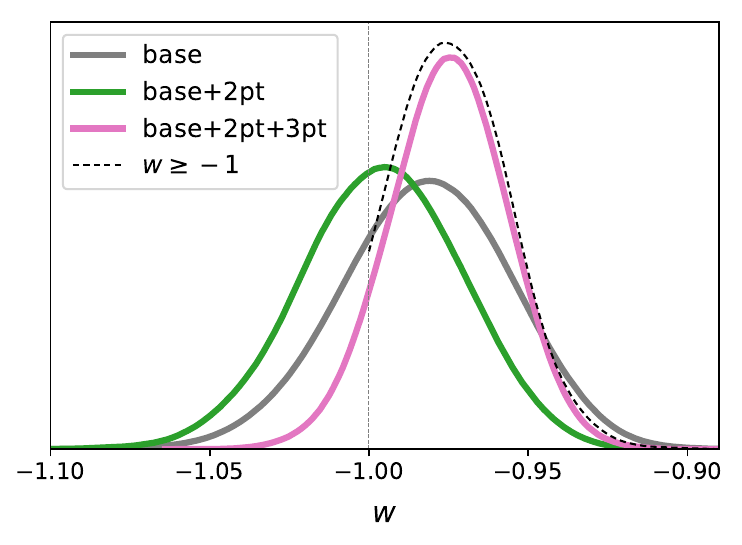}
\caption{Posterior distributions of the dark energy equation of state $w$ from our base combined probes and in combination with BOSS 1-loop 2pt + 3pt. 
The dashed line corresponds to the full combination, `base+2pt+3pt', but restricting $w \geq -1$.}
\label{fig:wcdm_1d_plot}
\end{figure}

\begin{figure}[ht]
\includegraphics[width=0.48\textwidth]{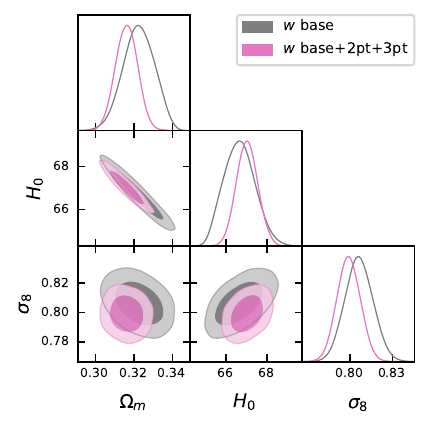}\\
\includegraphics[width=0.48\textwidth]{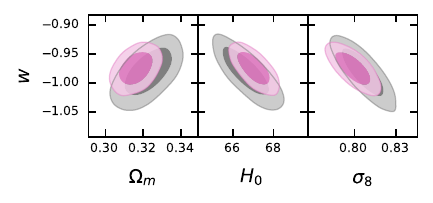}
\caption{$w$CDM triangle plot of $w$, $\Omega_m$, $H_0$, and $\sigma_8$ from our base combined probes and in combination with BOSS 1-loop 2pt + 3pt. }
\label{fig:wcdm_triangle_plot_small}
\end{figure}

\begin{table*}[]
    \centering
    \begin{tabular}{|l|c|c|c|c|}
    \hline
    \multicolumn{5}{|c|}{$w$CDM} \\
    \hline
        & base &  base+BAO & base+2pt &  base+2pt+3pt \\
    \hline
$w$
        & $-0.982\pm 0.027$
        & $-0.987\pm 0.026$
        & $-0.996\pm 0.026$
        & $-0.975\pm 0.019$
    \\
$10^{2}\omega_b$
        & $2.233\pm 0.014$
        & $2.239\pm 0.014$
        & $2.236\pm 0.014$
        & $2.241\pm 0.014$
    \\
$\omega_{\rm cdm}$
        & $0.1203\pm 0.0012$
        & $0.1194\pm 0.0011$
        & $0.1199\pm 0.0010$
        & $0.1190\pm 0.0010$
    \\
$H_0 \ [\textrm{km/s/Mpc}]$
        & $66.67^{+0.70}_{-0.80}$
        & $67.20\pm 0.66$
        & $67.28\pm 0.69$
        & $67.03\pm 0.51$
    \\
$\ln(10^{10}A_s)$
        & $3.046\pm 0.014$
        & $3.049\pm 0.015$
        & $3.047\pm 0.014$
        & $3.042\pm 0.014$
    \\
$n_s$
        & $0.9635\pm 0.0040$
        & $0.9658\pm 0.0039$
        & $0.9644\pm 0.0040$
        & $0.9665\pm 0.0038$
    \\
$\tau_{\rm reio}$
        & $0.0548\pm 0.0074$
        & $0.0573\pm 0.0076$
        & $0.0557\pm 0.0072$
        & $0.0545\pm 0.0071$
    \\
    \hline
$\Omega_m$
        & $0.3225\pm 0.0081$
        & $0.3155\pm 0.0067$
        & $0.3158\pm 0.0070$
        & $0.3162\pm 0.0057$
    \\
$\sigma_8$
        & $0.8064\pm 0.0098$
        & $0.8066\pm 0.0098$
        & $0.8096\pm 0.0096$
        & $0.7991\pm 0.0082$
    \\
$S_8$
        & $0.836\pm 0.012$
        & $0.827\pm 0.011$
        & $0.831\pm 0.011$
        & $0.820\pm 0.011$
    \\
    \hline
    \end{tabular}
    \caption{$w$CDM $68\%$ credible intervals from the various dataset combinations considered in this work.  }
    \label{tab:wcdm}
\end{table*}

We now present our $w$CDM results obtained from the various dataset combinations considered in this work. 
In Fig.~\ref{fig:wcdm_triangle_plot_small}, we show the triangle plot of $w, \Omega_m, H_0$, and $\sigma_8$ and in Fig.~\ref{fig:wcdm_1d_plot} the 1D posterior distributions of $w$, while the full cosmological triangle plot is provided in App.~\ref{app:triangle_plots}. 
The corresponding $68\%$ credible intervals are given in Tab.~\ref{tab:wcdm}. 
First, we note that for all cosmological parameters, the constraints from the various combinations are consistent within $1\sigma$. 
Albeit minor differences in dataset selection, our `base+BAO' results are consistent with those presented in the Pantheon+ collaboration multi-probe analysis~\cite{Brout:2022vxf}. 
Our full combination including the galaxy 1-loop 2pt+3pt reduces the uncertainties on $w, \Omega_m$, and $H_0$ by about $30-32\%$ with respect to our baseline multi-probe, and respectively $28\%, 15\%$, and $23\%$ with respect to `base+BAO'. 
Notably, the correlation between $\Omega_m$ and $w$ decreases from $\sim 0.7$ in `base+2pt' to $0.45$ `base+2pt+3pt' (see Fig.~\ref{fig:wcdm_triangle_plot_small}), showing that the 1-loop 3pt helps break parameter degeneracies. 
This reduction of $\sim 35\%$ in the $\Omega_m - w$ correlation translates to a roughly $20\%$ reduction on $w$ when marginalizing over $\Omega_m$. 
This improvement stands in contrast with the combination that includes only the galaxy 1-loop 2pt, which reduces uncertainties on $w, \Omega_m$, and $H_0$ by a limited amont of about $4\%, 15\%$, and $10\%$, respectively, compared to the baseline multi-probe or `base+BAO'. 
When including the BOSS bispectrum but analyzed at tree-level, we obtain $w = -1.01\pm 0.025, \Omega_m = 0.3121\pm 0.0070, H_0 = 67.65\pm 0.68 \ \text{km/s/Mpc}$, and $\sigma_8 = 0.8118\pm 0.0093$ at $68\%$CL, with practically no improvement compared to the analysis that includes only the power spectrum. 
These results highlight the cosmological significance of higher modes from higher $N$-point functions, particularly in constraining deviations from $\Lambda$CDM. 

To gauge the potential impact of the prior on the EFT parameters described in Sec.~\ref{sec:likelihood} when sampling from the galaxy 1-loop 2pt+3pt likelihood, we present results obtained from `base+2pt+3pt' where we have doubled the range of the allowed region for the EFT parameters, compared to the one used in our main analyses. 
In this case, we obtain $w = -0.970\pm 0.020, \Omega_m = 0.3172\pm 0.0058, H_0 = 66.91\pm 0.53 \ \text{km/s/Mpc}$, and $\sigma_8 = 0.7967\pm 0.0086$ at $68\%$CL. 
These results represent reductions of less than $5\%$ in the determination of the cosmological parameters, and relative shifts in the mean values of less than a quarter. 
We conclude that our prior on the EFT parameters, chosen such as our predictions are kept within physical range, is marginally informative. 

As for the mean value of $w$, we observe that all combinations produce values consistent with $-1$, with the farthest deviation found in our full combination being at a $\sim 1.3\sigma$ level. 
Since the update in the SNIa catalog from Pantheon+~\cite{Brout:2022vxf}, the multi-probe analysis now favors values that are $ \gtrsim -1$, in contrast with previous SNIa catalog from Pantheon~\cite{Pan-STARRS1:2017jku} (see e.g.,~\cite{Planck:2018vyg,eBOSS:2020yzd}). 
It is worth noting that the inclusion of BOSS 1-loop 2pt+3pt shifts $w$ in the same direction as the SNIa update, reinforcing the observed trend. 
Still, at this stage, all combinations considered here yield results that do not reveal any significant deviation from the cosmological constant. 

From the point of view of the effective field theory, $w < -1$ is generally deemed unphysical due to the presence of ghosts, i.e., degrees of freedom with negative kinetic terms~\cite{Caldwell:1999ew,Cline:2003gs}. 
However, there are exceptions where $w < -1$ is allowed (see e.g.,~\cite{Carroll:2003st,Creminelli:2006xe,Creminelli:2008wc,Cai:2009zp,Deffayet:2021nnt}). 
In such cases, fluctuations in dark energy are expected to contribute to the gravitational potential, which, in principle, should be incorporated into predictions for galaxy statistics~\cite{Lewandowski:2016yce,Cusin:2017wjg,DAmico:2020tty}. 
The $w$CDM analyses in this work consider modifications from dark energy at the background level and at linear level in perturbations. 
As such, unless some UV mechanism prevents instability, strictly speaking, $w\geq-1$ should be the only physically allowed region. 
Still, taking a data-driven approach, in first approximation modifications at the nonlinear level can be neglected, thus motivating us to scan $w$ without boundaries. 
In fact, for clustering quintessence where $w < -1$ is allowed~\cite{Creminelli:2006xe,Creminelli:2008wc}, it was shown in Ref.~\cite{DAmico:2020tty} that neglecting modifications in the perturbations was shifting $w$ by a relatively small amount for BOSS, on the order of $\lesssim 0.3\sigma$. 
In Fig.~\ref{fig:wcdm_1d_plot}, we also present the posterior of $w$ when restricting $w \geq -1$. 
Given that the posterior mean falls within this region when incorporating BOSS 1-loop 2pt+3pt, the results are reasonably consistent with those allowing $w$ to vary freely. 
When restricting $w \geq -1$, we obtain $w = -0.963 \pm 0.017$~\footnote{We stress that this bound is reported for the accessible region. As it is clear from Fig.~\ref{fig:wcdm_1d_plot} and Tab.~\ref{tab:wcdm}, the actual credible interval is wider if the phantom region is included.  }, $\Omega_m = 0.3218 \pm 0.0060$, $H_0 = 66.51 \pm 0.51 \ \text{km/s/Mpc}$, and $\sigma_8=0.7972 \pm 0.0074$ at $68\%$CL. 
Looking ahead, it will be interesting to analyze $w$ within a broader context of dark energy and modified gravity (see e.g.,~\cite{Gubitosi:2012hu,Gleyzes:2014qga,Cusin:2017wjg,Piga:2022mge}).

\section{\label{sec:conclusion} Conclusions}

In this work, we have presented cosmological constraints from multi-probe with BOSS galaxy power spectrum and bispectrum at one-loop precision. 
Our analysis demonstrates that the inclusion of BOSS one-loop three-point statistics significantly enhances our constraining capabilities. 
For $\Lambda$CDM parameters, this results in a reduction of uncertainties by approximately $5-10\%$.  
Our inference leads to $\Omega_m=0.3135 \pm 0.0051$, $H_0=67.47\pm 0.38 \ \text{km/s/Mpc}$, and $\sigma_8=0.8063 \pm 0.0056$ at $68\%$CL, to about $1.62\%$, $0.56\%$, and $0.69\%$ precision, respectively. 
When letting the dark energy equation of state vary, the addition of BOSS one-loop bispectrum leads to the stringent limit of $w=-0.975 \pm 0.019$ at $68\%$CL, reducing over former analyses the uncertainties on $w$, $\Omega_m$, and $H_0$ by about $30\%$, $15\%$, and $25\%$ respectively. 
Remarkably, these precision improvements are achieved using the same set of experiments. 

As a final remark, shot noise in experiments becomes increasingly dominant at higher modes. 
For instance, with the expansion up to fifth order in perturbations~\cite{Schmidt:2020ovm,Donath:2023sav}, it might turn out that additional information will come mainly from the one-loop trispectrum rather than the two-loop power spectrum. 
We hope to explore these exciting possibilities in the near future. 

Our findings call for a systematic inclusion of higher-$N$-point functions at the highest reachable scales to optimally extract cosmological information from galaxy maps. 
As shown on Fisher matrix, the bispectrum of galaxies will play a decisive role in the next decade~\cite{Braganca:2023pcp,Cabass:2022epm}.

\section*{Acknowledgements}
We thank Leonardo Senatore for encouraging us to work on this project, as well as Guido d'Amico, Matt Lewandowski, and Vivian Poulin for insightful comments.

\bibliographystyle{JHEP}
\bibliography{mybib}

\clearpage
\newpage
\appendix
\onecolumngrid

\section{\label{app:triangle_plots} Full cosmology triangle plots}

\begin{figure}[ht]
\includegraphics[width=0.83\columnwidth]{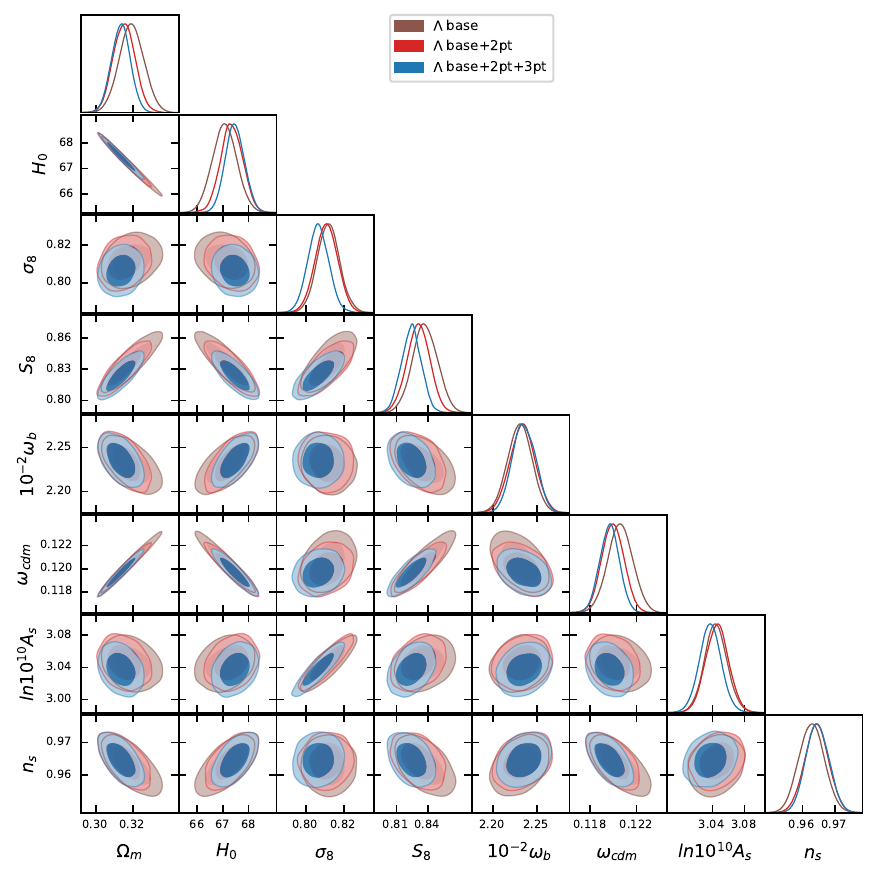}
\caption{$\Lambda$CDM triangle plots of cosmological parameters from our base combined probes and in combination with BOSS 1-loop 2pt + 3pt. }
\label{fig:lcdm_triangle_plot_full}
\end{figure}

\begin{figure}[ht]
\includegraphics[width=1.\columnwidth]{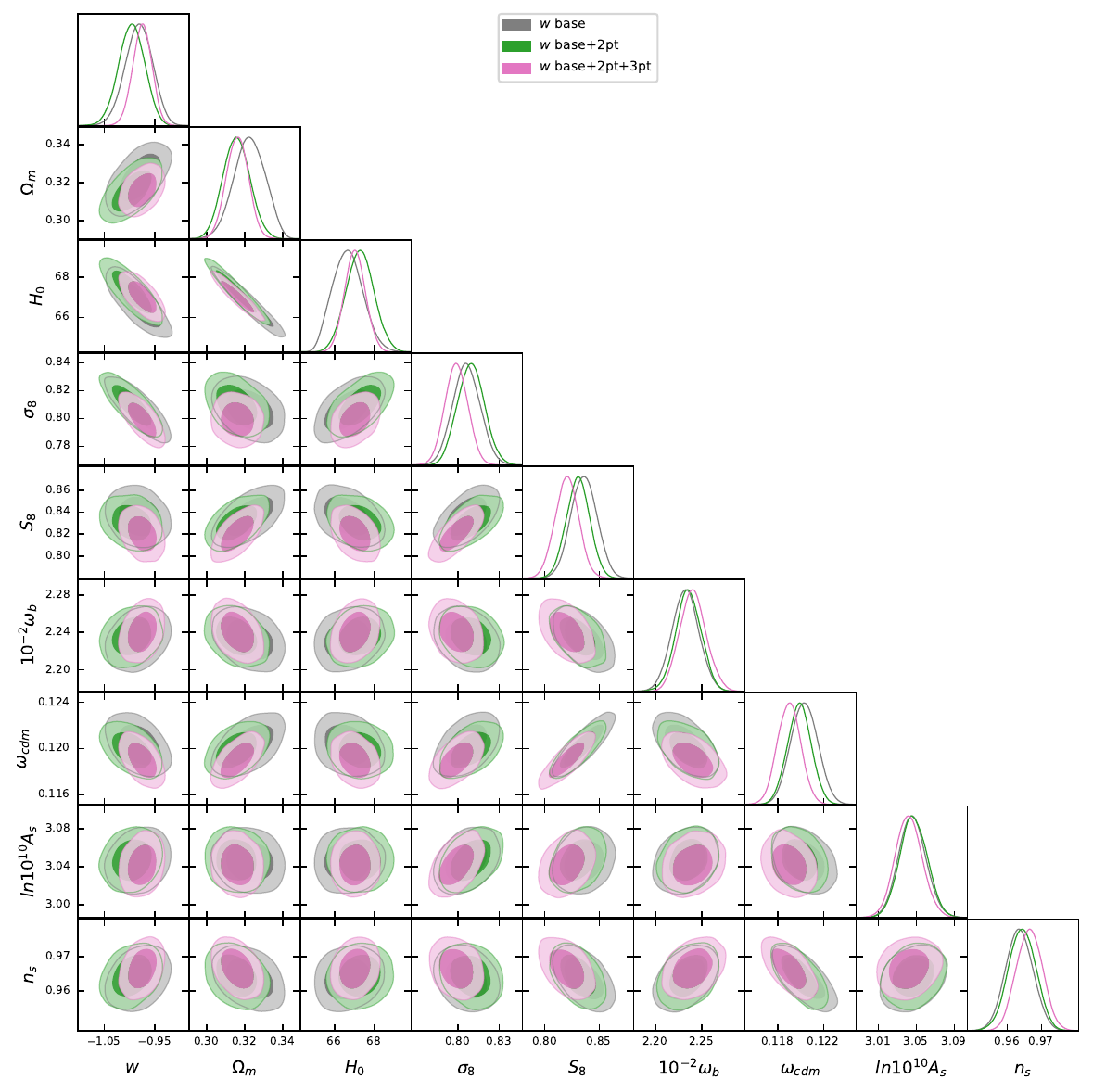}
\caption{$w$CDM triangle plots of cosmological parameters from our base combined probes and in combination with BOSS 1-loop 2pt + 3pt. }
\label{fig:wcdm_triangle_plot_full}
\end{figure}

In Figs.~\ref{fig:lcdm_triangle_plot_full}~and~\ref{fig:wcdm_triangle_plot_full}, we provide the full cosmology triangle plots of the posteriors from the $\Lambda$CDM and $w$CDM analyses carried in this work, respectively. 
 
\end{document}